\documentclass[12pt,doublespacing]{article}
\usepackage{fullpage}
\usepackage{setspace}
\usepackage{authblk}
\usepackage{cite}
\usepackage[title]{appendix}
\usepackage{amssymb}
\usepackage{graphicx}
\usepackage{graphics}
\usepackage[cmex10]{amsmath}
\usepackage{latexsym,amsfonts}
\usepackage{amsthm}
\usepackage{subfigure}
\usepackage{hyperref} 
\usepackage{enumitem}

\usepackage{booktabs}
\usepackage{makecell}

\begin{document}

\title{Sub-structure Characteristic Mode Analysis of Microstrip Antennas Using a Global Multi-trace Formulation}

\author[1]{Ran Zhao}
\author[2]{Yuyu Lu}
\author[2]{Guang Shang Cheng}
\author[3]{Wei Zhu}
\author[3]{Jun Hu}
\author[4]{Hakan Bagci}

\affil[1]{Ran Zhao is with the School of Electronic Science and Engineering, University of Electronic Science and Technology of China (UESTC), Chengdu 611731, China and also with the Electrical and Computer Engineering Program, Computer, Electrical, and Mathematical Science and Engineering Division, King Abdullah University of Science and Technology, Thuwal 23955, Saudi Arabia (e-mail: ran.zhao@kaust.edu.sa).}

\affil[2]{Yuyu Lu and Guang Shang Cheng are with the Information Materials and Intelligent Sensing Laboratory of Anhui Province, Anhui University, Hefei, China, 230601, and also with the Anhui Province Key Laboratory of Target Recognition and Feature Extraction, Lu'an, China, 237010  (email: 157064746@qq.com, gscheng89@ahu.edu.cn).}

\affil[3] {Wei Zhu and Jun Hu are with the School of Electronic Science and Engineering, University of Electronic Science and Technology of China (UESTC), Chengdu 611731, China (email: happyweiwei@uestc.edu.cn, hujun@uestc.edu.cn).}

\affil[4] {Hakan Bagci is with the Electrical and Computer Engineering Program, Computer, Electrical, and Mathematical Science and Engineering Division, King Abdullah University of Science and Technology (KAUST), Thuwal 23955, Saudi Arabia (e-mail: hakan.bagci@kaust.edu.sa).}

\footnotetext[1]{This work was supported in part by NSFC under Grant 62231007, Grant 62271002, Grant 62371001 and in part by KAUST OSR under Award 2019-CRG8-4056.}
\date{}
\maketitle
\newpage

\begin{abstract}
A characteristic mode (CM) method that relies on a global multi-trace formulation (MTF) of surface integral equations is proposed to compute the modes and the resonance frequencies of microstrip patch antennas with finite dielectric substrates and ground planes. Compared to the coupled formulation of electric field and Poggio-Miller-Chang-Harrington-Wu-Tsai integral equations, global MTF allows for more direct implementation of a sub-structure CM method. This is achieved by representing the coupling of the electromagnetic fields on the substrate and ground plane in the form of a numerical Green function matrix, which yields a more compact generalized eigenvalue equation. The resulting sub-structure CM method avoids the cumbersome computation of the multilayered medium Green function (unlike the CM methods that rely on mixed-potential integral equations) and the volumetric discretization of the substrate (unlike the CM methods that rely on volume-surface integral equations), and numerical results show that it is a reliable and accurate
approach to predicting the modal behavior of electromagnetic fields on practical microstrip antennas.
\end{abstract}
\newpage

\section{Introduction}
Atypical microstrip antenna consists of a metallic patch, a dielectric substrate that supports the patch, and a ground plane which the dielectric substrate is mounted to. Microstrip antennas and antenna arrays with microstrip elements are widely used in wireless systems since they are compact, lightweight, can be easily mounted on different types of surfaces, and their fabrication is relatively straightforward and often cheap~\cite{MSA1995}. Methods that can compute the modes and the resonance frequencies of a microstrip antenna are indispensable in the design process. To this end, several analytical methods have been developed, including transmission line model~\cite{TML1974,BPA2006} and cavity model~\cite{BPA2006,cavity1979} methods. However, it is difficult or even impossible to extend these analytical methods to account for antennas with irregular shapes or thick substrates. For such cases, characteristic mode analysis (CMA), which often relies on numerical modeling of an antenna, can be used to generate the modes and the resonance frequencies required by the design process~\cite{garbacz1982}.

CMA has first been applied to perfect electrically conducting (PEC) objects numerically modeled using scattering matrices~\cite{CM1971Garbacz} and surface integral equations (SIEs)~\cite{CM1971Harrington}. Later, it has been extended to account for dielectric and composite objects that are modeled using SIEs or volume integral equations (VIEs)~\cite{VIEdie1972, SIEdie1977, lian2017alternative, hu2016integral, guo2017characteristic, yla2019generalized, kuosmanen2022orthogonality}. These methods compute the characteristic modes (CMs) that are supported by the whole structure (which is accessible), and they are termed ``full-structure CM" methods in the literature. Another group of methods, which are developed to compute CMs and termed ``sub-structure CM" methods, first decompose the object into two regions (accessible feeding region and non-accessible coupling region), then reduce the generalized eigenvalue equation into a more compact form. This is done by re-expressing discretized electromagnetic field interactions in the non-accessible region as a ``numerical Green function matrix"~\cite{parhami1977technique,dai2016characteristic}. The sub-structure CM methods might be preferred in the design process of microstrip antennas since they directly provide the mode information supported by the radiation patch and the feeding region~\cite{huang2021accuratem}.

The idea of sub-structure CMA has first been developed for PEC objects~\cite{subJ2012}. Since then, it has been extended to account for a PEC object in the presence of a dielectric object, and vice-versa~\cite{subC2014}. Note that, the CM method, which relies on the mixed-potential integral equation (MPIE) to model microstrip patch antennas~\cite{CM2015, meng2018study}, can also be grouped as a sub-structure CM method, because the electromagnetic field interactions associated with the dielectric substrate and the metallic ground plane are accounted for using the multilayered medium Green function (under the assumption that the substrate and ground plane extend to infinity in transverse direction).

The orthogonality properties of the sub-structure CMs have been discussed in~\cite{alroughani2016orthogonality} and the differences between the full-structure and sub-structure CMs have been studied in~\cite{xiang2019preliminary} by replacing the substrate with air and simplifying the original composite objects into PEC objects. To compute the modes of practical microstrip patch antennas (with finite substrates and ground planes), a sub-structure CM method that relies on a volume-surface integral equation (VSIE) has been developed~\cite{VSIEwuqi}. It has been shown that this method leads to an antenna design with a better performance than that is obtained via the traditional full-structure CM methods~\cite{huang2021accuratem}. To avoid the high computational cost that comes with the volumetric discretization, a sub-structure CM method that relies only on SIEs has been developed to compute the modes of a dielectric resonator antenna with a finite ground plane~\cite{huang2021accurate}. However, this method calls for selection of a special weighting matrix to suppress the spurious (non-physical) modes that contaminate the actual CMs of the structure. Recently, a generalized SIE-based sub-structure CM method is proposed for arbitrary composite objects~\cite{huang2023generalized}, verifying the capability of the sub-structure CM method to analyze various kinds of practical antennas.

Another full-structure CM method that relies on the coupled electric field integral equation (EFIE) and Poggio-Miller-Chang-Harrington-Wu-Tsai (PMCHWT) equation~\cite{zhao2010analysis} has been developed in~\cite{fan2022spurious}. However, it is not trivial to transform this method into a sub-structure CM method. This is because the equivalent surface currents introduced on the metallic and the dielectric surfaces are not naturally separated. Fortunately, the global multi-trace formulation (MTF) developed in~\cite{lasisi2022fast} to analyze electromagnetic scattering from a dielectric object partially covered with a metallic sheet, can be adopted to alleviate this bottleneck. This method introduces a virtual gap to separate the adjacent metallic and dielectric domains.

In this work, first, global MTF presented in~\cite{lasisi2022fast} is carefully derived and it is used to develop a sub-structure CM equation to numerically characterize the modes and the resonance frequencies of practical microstrip patch antennas with finite substrates and ground planes. Note that in this formulation, radiation patch and its feeding portion are the accessible region (where the modes are computed) and the substrate and the ground plane are the non-accessible region. Having said that, the effects of the electromagnetic field interactions on the non-accessible region are fully accounted for, which can be interpreted as constructing and using a numerical Green function. Numerical results demonstrate that the proposed method can accurately predict the modal behavior of the electromagnetic fields and the associated resonance frequencies on practical microstrip antennas.

\section{Formulation}
\subsection{Global MTF}
\begin{figure}[t]
\centering
\subfigure[]{\includegraphics[width=0.49\columnwidth,draft=false]{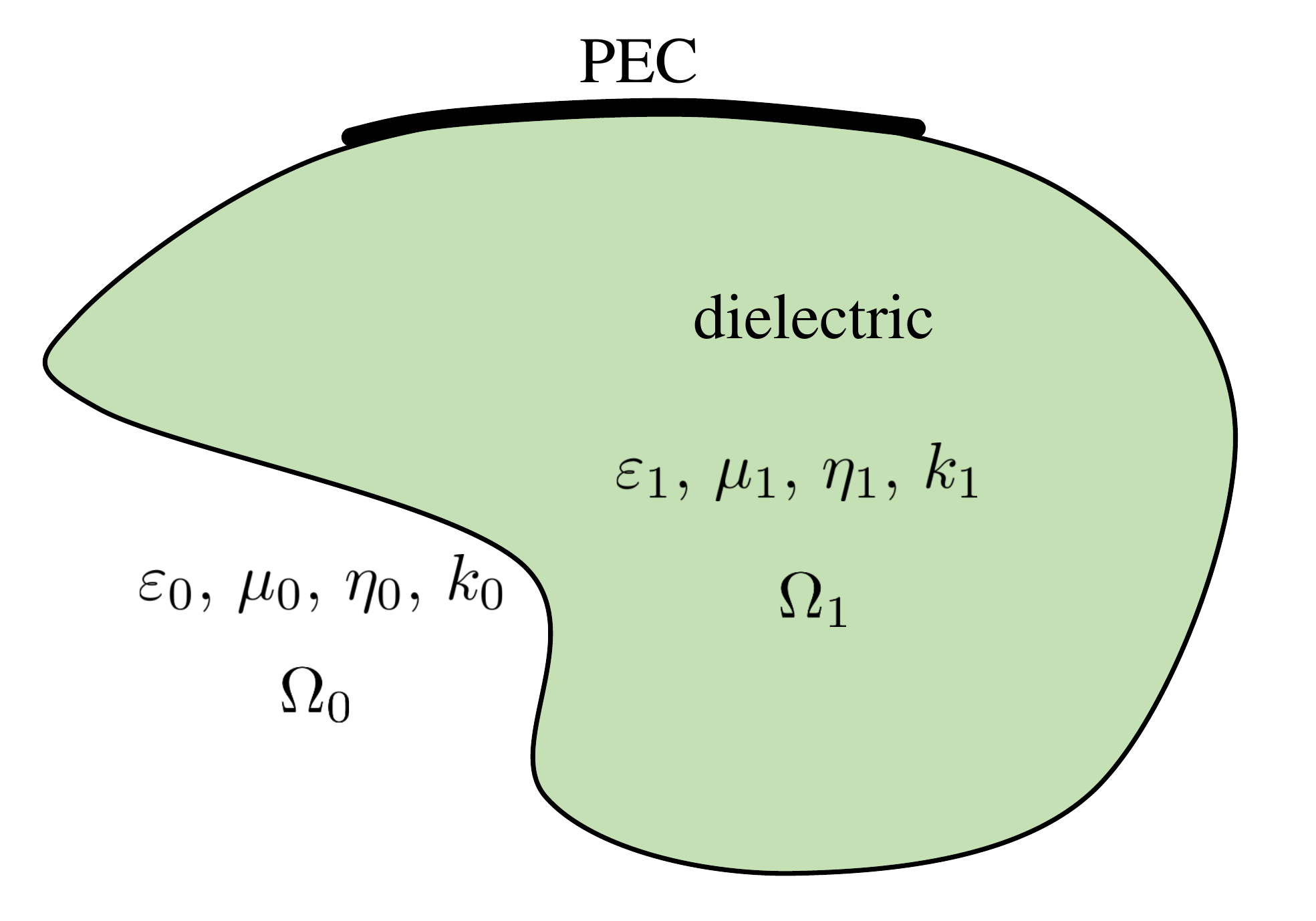}}
\subfigure[]{\includegraphics[width=0.49\columnwidth,draft=false]{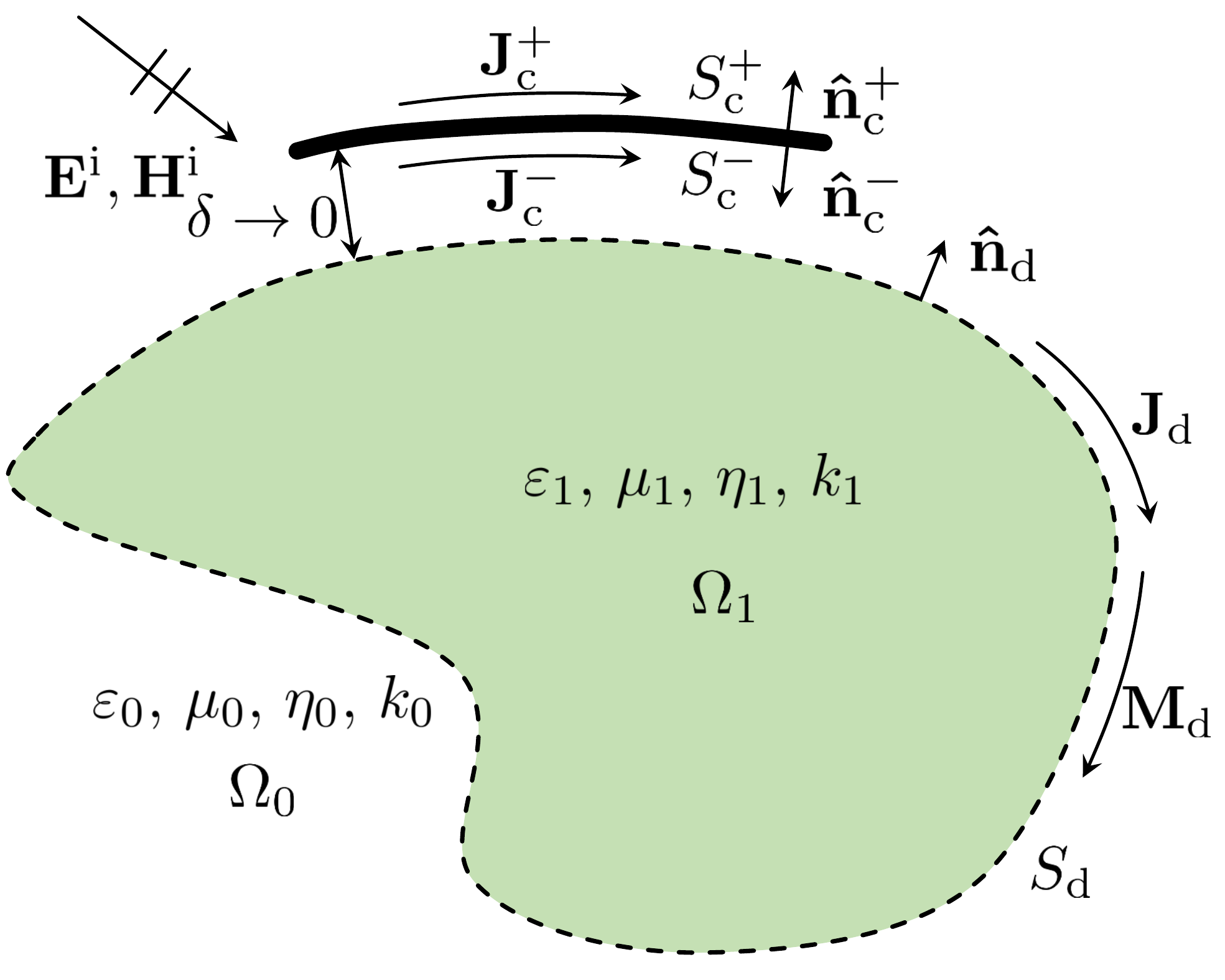}}
\caption{(a) Dielectric object partially covered by a PEC sheet. (b) Equivalent problems.}
\label{fig:patstucture}
\end{figure}

Consider a lossless dielectric object partially covered by an infinitesimally thin PEC sheet as shown in Fig.~\ref{fig:patstucture}(a). This composite structure resides in an unbounded background medium and is excited by an incident electromagnetic field $\{\mathbf{E}^\mathrm{i},\mathbf{H}^\mathrm{i}\}$. Let the background medium and the dielectric region be represented by $\Omega _0$ and $\Omega _1$, respectively. The permittivity, the permeability, the wave impedance, and the wave number in $\Omega _m, m = 0, 1,$ are denoted by $\varepsilon _m$, $\mu _m$, ${\eta _m}$, and $k_m$, respectively. Let $S_{\mathrm c}$ and $S_{\mathrm d}$ represent the open surface of the PEC sheet and the closed surface of the dielectric object, respectively.

Surfaces $S_{\mathrm c}$ and $S_{\mathrm d}$ are ``seperated'' from each other by introducing an infinitesimally thin \emph{virtual} gap (same medium as the background) as shown in Fig.~\ref{fig:patstucture}(b). Different from the formulation described in~\cite{lasisi2022fast}, two sets of equivalent electric currents $\mathbf{J}_{\mathrm{c}}^+(\mathbf{r})$ and $\mathbf{J}_{\mathrm{c}}^-(\mathbf{r})$ are defined on the upper side $S^{\mathrm{+}}_{\mathrm{c}}$ (the side that ``touches'' the background medium in Fig.~\ref{fig:patstucture}(b)) and the lower side $S^{\mathrm{-}}_{\mathrm{c}}$ (the side that ``touches'' the dielectric object in Fig.~\ref{fig:patstucture}(b)) of $S_{\mathrm c}$, respectively. Equivalent electric current $\mathbf{J}_\mathrm{d}(\mathbf{r})$ and the equivalent magnetic current $\mathbf{M}_\mathrm{d}(\mathbf{r})$ are defined on $S_\mathrm{d}$. In the exterior and interior equivalent problems, $S^{\mathrm{+}}_{\mathrm{d}}$ and $S^{\mathrm{-}}_{\mathrm{d}}$ represent the ``outer'' and the ``inner'' sides of $S_{\mathrm d}$, respectively. In Fig.~\ref{fig:patstucture}(b), $\mathbf{\hat{n}}_{\mathrm{d}}(\mathbf{r})$, $\mathbf{\hat{n}}_{\mathrm{c}}^+(\mathbf{r})$, and $\mathbf{\hat{n}}_{\mathrm{c}}^-(\mathbf{r})$ denote the outward pointing unit normal vectors of $S_\mathrm{d}$, $S^{\mathrm{+}}_{\mathrm{c}}$, and $S^{\mathrm{-}}_{\mathrm{c}}$, respectively. 

Let $\mathcal{L}_m\{\mathbf{X}\}(\mathbf{r})$ and $\mathcal{K}_m\{\mathbf{X}\}(\mathbf{r})$ represent the integral operators defined as~\cite{fan2022spurious}
\begin{equation}
\label{eq:1a} {\mathcal{L}_m}\{ \mathbf{X}\}(\mathbf{r})
= - j{k_m}\int_{S} {\left( {\mathbf{I} + \frac{1}{{k_m^2}}\nabla \nabla  \cdot } \right){G_m}( {\mathbf{r},\mathbf{r}'} )\mathbf{X}( {\mathbf{r}'} )d\mathbf{r}'}
\end{equation}
\begin{equation}
\begin{aligned}
\label{eq:1b} {\mathcal{K}_m}\{ \mathbf{X}\}(\mathbf{r}) & = \int_{S} {\nabla {G_m}( {\mathbf{r},\mathbf{r}'} ) \times \mathbf{X}( {\mathbf{r}'} )d\mathbf{r}'} \\
&=\pm\frac{1}{2}\mathbf{X}( {\mathbf{r}} ) \times {{\mathbf{\hat{n}}}}( {\mathbf{r}} ) + {\mathcal{\overline{\mathcal{K}}}_m}\{ \mathbf{X} \}(\mathbf{r})
\end{aligned}
\end{equation}
where $G_m(\mathbf{r}, \mathbf{r}^{\prime})=e^{-j k_m|\mathbf{r}-\mathbf{r}^{\prime}|} /(4 \pi|\mathbf{r}-\mathbf{r}^{\prime}|)$ is the Green function of the unbounded medium with permittivity $\varepsilon_m$ and permeability $\mu_m$ and $S$ represents the surface on which $\mathbf{X}$ is defined. In~\eqref{eq:1b}, $\hat{\mathbf{n}}(\mathbf{r})$ is the outward pointing unit normal vector of $S$, ${\overline{\mathcal{K}}}_m\{\mathbf{X}\} (\mathbf{r})$ represents the principle value integral term of ${\mathcal{K}_m}\{ \mathbf{X}\}(\mathbf{r})$, and `$-$' and `$+$' signs (in front of the residue term) are selected when $\mathbf{r}$ approaches $S$ along the direction and the opposite direction of $\hat{\mathbf{n}}(\mathbf{r})$, respectively. Note that in the remainder of the paper, the dependence of the variables on $\mathbf{r}$ is dropped for the sake of simplicity in the notation.

For the exterior equivalent problem, electromagnetic fields and the equivalent currents satisfy
\begin{equation}
\begin{aligned}
&\mathbf{M}_{\mathrm{c}}^{\mathrm{+}}  =\frac{1}{\eta_0} ( {{\mathbf{E}^{\mathrm{i}}} + {\mathbf{E}}_0^{\mathrm{s}}} ) \times {\mathbf{\hat{n}}_{\mathrm{c}}^{\mathrm{+}}}= 0 ,\mathbf{r} \in {S^{\mathrm{+}}_{\mathrm{c}}}\\
&\mathbf{M}_{\mathrm{c}}^{\mathrm{-}}  = \frac{1}{\eta_0} ( {{\mathbf{E}^{\mathrm{i}}} + {\mathbf{E}}_0^{\mathrm{s}}} ) \times {\mathbf{\hat{n}}_{\mathrm{c}}^{\mathrm{-}}} = 0,\mathbf{r} \in {S^{\mathrm{-}}_{\mathrm{c}}}\\
&{{\mathbf{M}}_\mathrm{d}} = \frac{1}{\eta_0} ( {{\mathbf{E}^{\mathrm{i}}} + {\mathbf{E}}_0^{\mathrm{s}}} ) \times {\mathbf{\hat{n}}_{\mathrm{d}}},\mathbf{r} \in {S^{\mathrm{+}}_{\mathrm{d}}}\\
&{{\mathbf{J}_{\mathrm{d}}} = {\mathbf{\hat{n}}_{\mathrm{d}}} \times ( {{\mathbf{H}^{\mathrm{i}}} + \mathbf{H}_0^{\mathrm{s}}} ),\mathbf{r}\in {S^{\mathrm{+}}_{\mathrm{d}}}}.
\end{aligned}
\label{eq:2}
\end{equation}
Here, $\mathbf{E}_0^{\mathrm{s}}$ and $\mathbf{H}_0^{\mathrm{s}}$ are the scattered electric and magnetic fields in $\Omega_0$ and expressed in terms of the equivalent currents using~\cite{{lasisi2022fast}}
\begin{equation}
\begin{aligned}
\!\!{\mathbf{E}}_0^{\mathrm{s}} & = {\eta_0}\big(\mathcal{L}_0\{{\mathbf{J_{\mathrm{c}}^+}}\}
\!+\!{\mathcal{L}_0}\{{\mathbf{J_{\mathrm{c}}^-}}\}\!+\!{\mathcal{L}_0}\{{\mathbf{J_{\mathrm{d}}}}\}\!-\!{\mathcal{K}_0}\{ {\mathbf{M}_{\mathrm{d}}}\}\big),\mathbf{r} \in {\Omega _0}\\
\!\!\mathbf{H}_0^{\mathrm s} & =\mathcal{K}_0\{\mathbf{J}_{\mathrm c}^{+}\}\!+\!\mathcal{K}_0\{\mathbf{J}_{\mathrm c}^{-}\}\!+\!\mathcal{K}_0\{\mathbf{J}_{\mathrm d}\}\!+\!\mathcal{L}_0\{\mathbf{M}_{\mathrm d}\}, \mathbf{r} \in \Omega_0.
\end{aligned}
\label{eq:3}
\end{equation}
Using~\eqref{eq:2} and~\eqref{eq:3}, and the fact that $\hat{\mathbf{n}}_{\mathrm{c}}^{\pm}\times (\mathbf{M}_{\mathrm{d}} \times \hat{\mathbf{n}}_{\mathrm{d}})\times \hat{\mathbf{n}}_{\mathrm{c}}^{\pm}= \mathbf{M}_{\mathrm{d}} \times \hat{\mathbf{n}}_{\mathrm{d}}$, $\hat{\mathbf{n}}_{\mathrm{d}} \times (\mathbf{J}_{\mathrm{c}}^{{\pm}} \times \hat{\mathbf{n}}_{\mathrm{c}}^{{\pm}})\times \hat{\mathbf{n}}_{\mathrm{d}} = \mathbf{J}_{\mathrm{c}}^{{\pm}} \times \hat{\mathbf{n}}_{\mathrm{c}}^{{\pm}}$, $\mathbf{r} \in S_{\mathrm c}^{\pm}$, EFIEs on $S_{\mathrm{c}}^{+}$, $S_{\mathrm{c}}^{-}$, and $S_{\mathrm{d}}^{+}$ and the magnetic field integral equation (MFIE) on $S_{\mathrm{d}}^{+}$ can be obtained for the exterior problem as
\begin{equation}
\begin{aligned}
& \eta_0 \frac{\mathbf{M}_{\mathrm{d}}}{2} \times \hat{\mathbf{n}}_{\mathrm{d}}-\eta_0 \hat{\mathbf{n}}_{\mathrm{c}}^{+} \times\big(\mathcal{L}_0\{\mathbf{J}_{\mathrm{c}}^{+}\}+\mathcal{L}_0\{\mathbf{J}_{\mathrm{c}}^{-}\}\big.\\
&\big.+\mathcal{L}_0\{\mathbf{J}_{\mathrm{d}}\}-\overline{\mathcal{K}}_0\{\mathbf{M}_{\mathrm{d}}\}\big) \times \hat{\mathbf{n}}_{\mathrm{c}}^{+}=\hat{\mathbf{n}}_{\mathrm{c}}^{+} \times \mathbf{E}^{\mathrm{i}} \times \hat{\mathbf{n}}_{\mathrm{c}}^{+}, \mathbf{r} \in S_{\mathrm{c}}^{+}
\end{aligned}
\label{eq:4}
\end{equation}
\begin{equation}
\begin{aligned}
& \eta_0 \frac{\mathbf{M}_{\mathrm{d}}}{2} \times \hat{\mathbf{n}}_{\mathrm{d}}-\eta_0 \hat{\mathbf{n}}_{\mathrm{c}}^{-} \times\big(\mathcal{L}_0\{\mathbf{J}_{\mathrm{c}}^{+}\}+\mathcal{L}_0\{\mathbf{J}_{\mathrm{c}}^{-}\}\big.\\
&\big.+\mathcal{L}_0\{\mathbf{J}_{\mathrm{d}}\}-\overline{\mathcal{K}}_0\{\mathbf{M}_{\mathrm{d}}\}\big) \times \hat{\mathbf{n}}_{\mathrm{c}}^{-}=\hat{\mathbf{n}}_{\mathrm{c}}^{-} \times \mathbf{E}^{\mathrm{i}} \times \hat{\mathbf{n}}_{\mathrm{c}}^{-}, \mathbf{r} \in S_{\mathrm{c}}^{-}
\end{aligned}
\label{eq:5}
\end{equation}
\begin{equation}
\begin{aligned}
 & \eta_0 \hat{\mathbf{n}}_{\mathrm d} \times \frac{\mathbf{M}_{\mathrm{d}}}{2}-\eta_0 \hat{\mathbf{n}}_{\mathrm d} \times\big(\mathcal{L}_0\{\mathbf{J}_{\mathrm{c}}^{+}\}+\mathcal{L}_0\{\mathbf{J}_{\mathrm{c}}^{-}\}\big.\\
&\big.+\mathcal{L}_0\{\mathbf{J}_{\mathrm{d}}\}-\overline{\mathcal{K}}_0\{\mathbf{M}_{\mathrm{d}}\}\big) \times \hat{\mathbf{n}}_{\mathrm d}=\hat{\mathbf{n}}_{\mathrm d} \times \mathbf{E}^{\mathrm i} \times \hat{\mathbf{n}}_{\mathrm d}, \mathbf{r} \in S_{\mathrm{d}}^{+}
\end{aligned}
\label{eq:6}
\end{equation}
\begin{equation}
\begin{aligned}
& \eta_0 \frac{\mathbf{J}_{\mathrm{d}}}{2} \times \hat{\mathbf{n}}_{\mathrm{d}}+\eta_0\frac{\mathbf{J}_{\mathrm{c}}^{+}}{2} \times \hat{\mathbf{n}}_{\mathrm{c}}^{+}-\eta_0\frac{\mathbf{J}_{\mathrm{c}}^{-}}{2} \times \hat{\mathbf{n}}_{\mathrm{c}}^{-}\\
&-\eta_0 \hat{\mathbf{n}}_{\mathrm{d}} \times\big(\overline{\mathcal{K}}_0\left\{\mathbf{J}_{\mathrm{c}}^{+}\right\}+\overline{\mathcal{K}}_0\left\{\mathbf{J}_{\mathrm{c}}^{-}\right\}+\mathcal{L}_0\left\{\mathbf{M}_{\mathrm{d}}\right\}\big. \\
& \big.+\overline{\mathcal{K}}_0\left\{\mathbf{J}_{\mathrm{d}}\right\}\big) \times \hat{\mathbf{n}}_{\mathrm{d}}=\eta_0 \hat{\mathbf{n}}_{\mathrm{d}} \times \mathbf{H}^{\mathrm{i}} \times \hat{\mathbf{n}}_{\mathrm{d}}, \mathbf{r} \in S_{\mathrm{d}}^{+}.
\end{aligned}
\label{eq:7}
\end{equation}
Because $S_\mathrm{c}$ is an open surface and $\hat{\mathbf{n}}_{\mathrm{c}}^{-} \times \mathbf{A} \times \hat{\mathbf{n}}_{\mathrm{c}}^{-}=\hat{\mathbf{n}}_{\mathrm{c}}^{+} \times \mathbf{A} \times \hat{\mathbf{n}}_{\mathrm{c}}^{+}$ for any vector $\mathbf{A}$, \eqref{eq:4} and~\eqref{eq:5} are equivalent to each other and they can be merged into a single equation by defining a new electric current as the collection of $\mathbf{J}_{\mathrm{c}}^{+}$ and $\mathbf{J}_{\mathrm{c}}^{-}$, i.e., $\mathbf{J}_{\mathrm{c}} =\mathbf{J}_{\mathrm{c}}^{+}\cup \mathbf{J}_{\mathrm{c}}^{-}$. Noting that $\hat{\mathbf{n}}_{\mathrm{d}}=\pm\hat{\mathbf{n}}_{\mathrm{c}}^{\pm}, \mathbf{r} \in S_{\mathrm{c}}^{\pm}$, this new equation can be written as
\begin{equation}
\label{eq:8}
\begin{aligned}
& \eta_0 \frac{\mathbf{M}_{\mathrm{d}}}{2} \times \hat{\mathbf{n}}_{\mathrm{d}}-\eta_0 \hat{\mathbf{n}}_{\mathrm{d}} \times\big(\mathcal{L}_0\{\mathbf{J}_{\mathrm{c}}\}+\mathcal{L}_0\{\mathbf{J}_{\mathrm{d}}\}\big.\\
&\big.-\overline{\mathcal{K}}_0\{\mathbf{M}_{\mathrm{d}}\}\big) \times \hat{\mathbf{n}}_{\mathrm{d}}=\hat{\mathbf{n}}_{\mathrm{d}} \times \mathbf{E}^{\mathrm{i}} \times \hat{\mathbf{n}}_{\mathrm{d}}, \mathbf{r} \in S_{\mathrm{c}}.
\end{aligned}
\end{equation}
Using $\mathbf{J}_{\mathrm{c}} =\mathbf{J}_{\mathrm{c}}^{+}\cup \mathbf{J}_{\mathrm{c}}^{-}$ in~\eqref{eq:6} yields
\begin{equation}
\label{eq:9}
\begin{aligned}
& \eta_0\hat{\mathbf{n}}_{\mathrm{d}}\times\frac{\mathbf{M}_{\mathrm{d}}}{2}-\eta_0 \hat{\mathbf{n}}_{\mathrm{d}} \times\big(\mathcal{L}_0\{\mathbf{J}_{\mathrm{c}}\}+\mathcal{L}_0\{\mathbf{J}_{\mathrm{d}}\}\big.\\
&\big.-\overline{\mathcal{K}}_0\{\mathbf{M}_{\mathrm{d}}\}\big) \times \hat{\mathbf{n}}_{\mathrm{d}}=\hat{\mathbf{n}}_{\mathrm{d}} \times \mathbf{E}^{\mathrm{i}} \times \hat{\mathbf{n}}_{\mathrm{d}}, \mathbf{r} \in S_{\mathrm{d}}^{+}.
\end{aligned}
\end{equation}
Similarly, using $\mathbf{J}_{\mathrm{c}} =\mathbf{J}_{\mathrm{c}}^{+}\cup \mathbf{J}_{\mathrm{c}}^{-}$ and $\hat{\mathbf{n}}_{\mathrm{c}}^{+}=-\hat{\mathbf{n}}_{\mathrm{c}}^{-}$ in~\eqref{eq:7} yields
\begin{equation}
\label{eq:10}
\begin{aligned}
& \eta_0 \frac{\mathbf{J}_{\mathrm{d}}+\mathbf{J}_{\mathrm{c}}}{2} \times \hat{\mathbf{n}}_{\mathrm{d}}-\eta_0 \hat{\mathbf{n}}_{\mathrm{d}} \times\big(\overline{\mathcal{K}}_0\{\mathbf{J}_{\mathrm{c}}\}+\mathcal{L}_0\{\mathbf{M}_{\mathrm{d}}\}\big. \\
& \left.+\overline{\mathcal{K}}_0\{\mathbf{J}_{\mathrm{d}}\}\right) \times \hat{\mathbf{n}}_{\mathrm{d}}=\eta_0 \hat{\mathbf{n}}_{\mathrm{d}} \times \mathbf{H}^{\mathrm{i}} \times \hat{\mathbf{n}}_{\mathrm{d}}, \mathbf{r} \in S_{\mathrm{d}}^{+}.
\end{aligned}
\end{equation}
Note that $\hat{\mathbf{n}}_{\mathrm{d}}\times {{\mathbf{M}}_{\mathrm{d}}}=0$, $\mathbf{r} \in {S}_{\mathrm{c}}\cap{S}_{\mathrm{d}}^{+}$, since ${S}_{\mathrm{c}}$ is a PEC surface. The same boundary condition can be obtained by subtracting~\eqref{eq:8} from~\eqref{eq:9}.

For the interior equivalent problem, electromagnetic fields and the equivalent currents satisfy
\begin{align}
\begin{aligned}
& -\mathbf{M}_{\mathrm{d}}=\frac{1}{\eta_0}\mathbf{E}_1^{\mathrm s} \times(-\hat{\mathbf{n}}_{\mathrm d}), \mathbf{r} \in S_{\mathrm{d}}^{-} \\
& -\mathbf{J}_{\mathrm{d}}=(-\hat{\mathbf{n}}_{\mathrm d}) \times \mathbf{H}_1^{\mathrm s}, \mathbf{r} \in S_{\mathrm{d}}^{-}
\end{aligned}
\label{eq:11}
\end{align}
where $\mathbf{E}_1^{\mathrm{s}}$ and $\mathbf{H}_1^{\mathrm{s}}$ are the scattered electric and magnetic fields in $\Omega_1$ and expressed in terms of the equivalent currents using~\cite{lasisi2022fast}
\begin{equation}
\label{eq:scat1}
\begin{aligned}
& \mathbf{E}_1^{\mathrm s}=\eta_1 \mathcal{L}_1\{\mathbf{J}_{\mathrm d}\}-\eta_0 \mathcal{K}_1\{\mathbf{M}_{\mathrm d}\}, \mathbf{r} \in \Omega_1 \\
& \mathbf{H}_1^{\mathrm s}=\mathcal{K}_1\{\mathbf{J}_{\mathrm d}\}+\frac{\eta_0}{\eta_1} \mathcal{L}_1\{\mathbf{M}_{\mathrm d}\}, \mathbf{r} \in \Omega_1.
\end{aligned}
\end{equation}
Using~\eqref{eq:11} and~\eqref{eq:scat1}, one can obtain EFIE and MFIE on $S_{\mathrm d}^{-}$ for the interior problem as
\begin{equation}
\label{eq:12}
\begin{aligned}
\eta_0 \hat{\mathbf{n}}_{\mathrm d} \times \frac{\mathbf{M}_{\mathrm{d}}}{2}&+\hat{\mathbf{n}}_{\mathrm d} \times\big(\eta_1 \mathcal{L}_1\left\{\mathbf{J}_{\mathrm{d}}\right\}-\eta_0 \overline{\mathcal{K}}_1\left\{\mathbf{M}_{\mathrm{d}}\right\}\big) \times \hat{\mathbf{n}}_{\mathrm d} \\
& =0, \mathbf{r} \in S_{\mathrm{d}}^{-}
\end{aligned}
\end{equation}
\begin{equation}
\label{eq:13}
\begin{aligned}
\eta_0 \frac{\mathbf{J}_{\mathrm{d}}}{2}\times \hat{\mathbf{n}}_{\mathrm d}&+\hat{\mathbf{n}}_{\mathrm d} \times\big(\eta_0 \overline{\mathcal{K}}_1\left\{\mathbf{J}_{\mathrm{d}}\right\}+\frac{\eta_0^2}{\eta_1} \mathcal{L}_1\left\{\mathbf{M}_{\mathrm{d}}\right\}\big) \times \hat{\mathbf{n}}_{\mathrm d} \\
& =0, \mathbf{r} \in S_{\mathrm{d}}^{-}.
\end{aligned}
\end{equation}

Subtracting~\eqref{eq:12} from~\eqref{eq:9} and subtracting~\eqref{eq:13} from~\eqref{eq:10} yield two equations on $S_{\mathrm d}$ as
\begin{equation}
\begin{aligned}
& \hat{\mathbf{n}}_{\mathrm d} \times\big(-\eta_0 \mathcal{L}_0\{\mathbf{J}_{\mathrm{d}}\}-\eta_1 \mathcal{L}_1\{\mathbf{J}_{\mathrm{d}}\}+\eta_0 \overline{\mathcal{K}}_0\{\mathbf{M}_{\mathrm{d}}\}\big. \\
& \big.+\eta_0 \overline{\mathcal{K}}_1\{\mathbf{M}_{\mathrm{d}}\}-\eta_0 \mathcal{L}_0\{\mathbf{J}_{\mathrm{c}}\}\big) \times \hat{\mathbf{n}}_{\mathrm d}\\
&=\hat{\mathbf{n}}_{\mathrm d} \times \mathbf{E}^{\mathrm i} \times \hat{\mathbf{n}}_{\mathrm d}, \mathbf{r} \in S_{\mathrm{d}} \\
\end{aligned}
\label{eq:14}
\end{equation}
\begin{equation}
\begin{aligned}
& \eta_0 \frac{\mathbf{J}_{\mathrm{c}}}{2} \times \hat{\mathbf{n}}_{\mathrm d}+\hat{\mathbf{n}}_{\mathrm d} \times\big(-\eta_0 \overline{\mathcal{K}}_0\{\mathbf{J}_{\mathrm{d}}\}-\eta_0 \overline{\mathcal{K}}_1\{\mathbf{J}_{\mathrm{d}}\}\big.\\
& -\eta_0 \mathcal{L}_0\{\mathbf{M}_{\mathrm{d}}\}-\frac{\eta_0^2}{\eta_1} \mathcal{L}_1\{\mathbf{M}_{\mathrm{d}}\}-\eta_0 \overline{\mathcal{K}}_0\{\mathbf{J}_{\mathrm{c}}\}\big) \times \hat{\mathbf{n}}_{\mathrm d}\\
&=\eta_0 \hat{\mathbf{n}}_{\mathrm d} \times \mathbf{H}^{\mathrm i} \times \hat{\mathbf{n}}_{\mathrm d}, \mathbf{r} \in S_{\mathrm{d}}.
\end{aligned}
\label{eq:15}
\end{equation}

The system of surface integral equations~\eqref{eq:8},~\eqref{eq:14}, and~\eqref{eq:15} can be numerically solved for unknown equivalent currents $\mathbf{J}_{\mathrm{c}}$, $\mathbf{J}_{\mathrm{d}}$, and $\mathbf{M}_{\mathrm{d}}$. First, $S_{\mathrm c}$ and $S_{\mathrm d}$ are discretized into a mesh of triangles. Then, $\mathbf{J}_{\mathrm{c}}$, $\mathbf{J}_{\mathrm{d}}$, and $\mathbf{M}_{\mathrm{d}}$ are expanded using the well-known Rao-Wilton-Glisson (RWG) basis functions, each of which is defined on a pair of triangles. Inserting these expansion into~\eqref{eq:8},~\eqref{eq:14}, and~\eqref{eq:15} and applying Galerkin testing to resulting equations yield a matrix system as
\begin{equation}
\begin{aligned}
&\begin{bmatrix}
\eta_0 \bar{\bar{L}}_{\mathrm {dd}}^0+\eta_1 \bar{\bar{L}}_{\mathrm {dd}}^1 & -\eta_0 \bar{\bar{K}}_{\mathrm {dd}}^0-\eta_0 \bar{\bar{K}}_{\mathrm {dd}}^1 & \eta_0 \bar{\bar{L}}_{\mathrm {dc}}^0 \\
\eta_0 \bar{\bar{K}}_{\mathrm {dd}}^0+\eta_0 \bar{\bar{K}}_{\mathrm {dd}}^1 & \eta_0 \bar{\bar{L}}_{\mathrm {dd}}^0+\frac{\eta_0^2}{\eta_1} \bar{\bar{L}}_{\mathrm {dd}}^1 & \eta_0 \bar{\bar{K}}_{\mathrm {dc}}^0+\frac{\eta_0}{2} \bar{\bar{S}}_{\mathrm {dc}} \\
\eta_0 \bar{\bar{L}}_{\mathrm {cd}}^0 & -\eta_0 \bar{\bar{K}}_{\mathrm {cd}}^0+\frac{\eta_0}{2} \bar{\bar{S}}_{\mathrm {cd}} & \eta_0 \bar{\bar{L}}_{\mathrm {cc}}^0
\end{bmatrix}\\
&\begin{bmatrix}
\bar{J}_{\mathrm d}\\
\bar{M}_{\mathrm d}\\
\bar{J}_{\mathrm c}
\end{bmatrix} =
\begin{bmatrix}
\bar{E}_{\mathrm d}\\
\eta_0\bar{H}_{\mathrm d}\\
\bar{E}_{\mathrm c}
\end{bmatrix}.
\end{aligned}
\label{eq:16}
\end{equation}
Here, the unknown coefficients of the basis function sets used to expand $\mathbf{J}_{\mathrm{c}}$, $\mathbf{J}_{\mathrm{d}}$, and $\mathbf{M}_{\mathrm{d}}$ are stored in vectors $\bar{J}_{\mathrm{c}}$, $\bar{J}_{\mathrm{d}}$, and $\bar{M}_{\mathrm{d}}$, respectively. Vectors $\bar{E}_{\mathrm d}$, $\bar{H}_{\mathrm d}$, and $\bar{E}_{\mathrm c}$ store tested incident fields and matrix blocks $\bar{\bar{L}}^m_{\mathrm{ab}}$, $\bar{\bar{K}}^m_{\mathrm{ab}}$, and $\bar{\bar{S}}^m_{\mathrm{ab}}$, $\mathrm{a},\mathrm{b}\in \{\mathrm{c},\mathrm{d}\}$ store tested fields of the basis functions. Entries of these vectors and matrix blocks are given by
\begin{equation}
\begin{aligned}
& \{\bar{E}_{\mathrm{a}}\}_p=\langle\mathbf{f}^p_{\mathrm{a}}, \mathbf{E}^{\mathrm i}\rangle_{\mathrm{a}}\\
& \{\bar{H}_{\mathrm{a}}\}_p=\langle\mathbf{f}^p_{\mathrm{a}}, \mathbf{H}^{\mathrm i}\rangle_{\mathrm{a}}\\
& \{\bar{\bar{L}}_{\mathrm{ab}}^m\}_{p, q}=-\langle\mathbf{f}^p_{\mathrm{a}}, \mathcal{L}_m\{\mathbf{f}^q_{\mathrm{b}}\}\rangle_{\mathrm{a}} \\
& \{\bar{\bar{K}}_{\mathrm{ab}}^m\}_{p, q}=-\langle\mathbf{f}^p_{\mathrm{a}}, \overline{\mathcal{K}}_m\{\mathbf{f}^q_{\mathrm{b}}\}\rangle_{\mathrm{a}} \\
& \{\bar{\bar{S}}_{\mathrm{ab}}\}_{p, q}=\langle\mathbf{f}^p_{\mathrm{a}}, \mathbf{f}^q_{\mathrm{b}} \times \hat{\mathbf{n}}_{\mathrm d}\rangle_{\mathrm{a}}.
\end{aligned}
\label{eq:17}
\end{equation}
Here, subscripts $\mathrm{a},\mathrm{b}\in \{\mathrm{c},\mathrm{d}\}$ attached to a variable mean that the variable is defined on/for surface $S_{\mathrm a}$ or $S_{\mathrm b}$, the inner product of two vector fields $\mathbf{u}$ and $\mathbf{v}$ is defined as $\langle \mathbf{u},\mathbf{v}\rangle_{\mathrm a}=\int_{S_{\mathrm{a}}}{{\mathbf{u} \cdot \mathbf{v}}\,ds}$, and $\mathbf{f}^p_{\mathrm{a}}$ and $\mathbf{f}^q_{\mathrm{b}}$ represent the $p$th testing function and the $q$th basis function used to test fields on $S_{\mathrm a}$ and expand unknown currents on $S_{\mathrm b}$, respectively.

\subsection{Sub-structure CM Method}
To faciliate the sub-structure CMA of a microstrip antenna where the radiation patch is the accessible region, matrix system~\eqref{eq:16} is brought into a form where the basis expansions of $\mathbf{J}_{\mathrm c}$ on the radiation patch (denoted by surface $S_{\mathrm r}$) and the finite ground plane (denoted by surface $S_{\mathrm g}$) can clearly be identified:
\begin{equation}
\begin{bmatrix}
\bar{\bar{Z}}_{11} & \bar{\bar{Z}}_{\mathrm{12}} \\
\bar{\bar{Z}}_{21} & \bar{\bar{Z}}_{\mathrm{22}}
\end{bmatrix}\begin{bmatrix}
\bar{I}_{\mathrm{1}} \\
\bar{I}_{\mathrm{2}}
\end{bmatrix}=\begin{bmatrix}
\bar{V}_{\mathrm{1}} \\
\bar{V}_{\mathrm{2}}
\end{bmatrix}.
\label{eq:18}
\end{equation}
Here, subscripts $\mathrm{r}$, $\mathrm{g}$, and $\mathrm{d}$ attached to a variable mean that the variable is defined on/for surface $S_{\mathrm r}$ (radiation patch) or $S_{\mathrm g}$ (ground plane), and $S_{\mathrm d}$ (dielectric surface), respectively. In~\eqref{eq:18}, $\bar{I}_1=\big[\bar{J}_{\mathrm{d}} \;\, \bar{M}_{\mathrm{d}} \;\, \bar{J}_{\mathrm{g}}\big]^{\mathrm{T}}$ and $\bar{I}_2 = \bar{J}_{\mathrm{r}}$, where $\bar{J}_{\mathrm{g}}$ and $\bar{J}_{\mathrm{r}}$ store the basis function expansion coefficients of $\mathbf{J}_{\mathrm c}$ on $S_{\mathrm g}$ and $S_{\mathrm r}$, respectively. Similarly, $\bar{V}_1=\big[\bar{E}_{\mathrm{d}} \;\, \eta_0\bar{H}_{\mathrm{d}} \;\, \bar{E}_{\mathrm{g}}\big]^{\mathrm{T}}$ and $\bar{V}_2 = \bar{E}_{\mathrm{r}}$, where $\bar{E}_{\mathrm{g}}$ and $\bar{E}_{\mathrm{r}}$ store the incident fields (or fields due to another type of excitation) tested on $S_{\mathrm g}$ and $S_{\mathrm r}$, respectively. The matrix blocks $\bar{\bar{Z}}_{11}$, $\bar{\bar{Z}}_{12}$, $\bar{\bar{Z}}_{21}$, and $\bar{\bar{Z}}_{22}$ are given by
\begin{equation}
\begin{aligned}
& \bar{\bar{Z}}_{11}= \\
& \begin{bmatrix}
\eta_0 \bar{\bar{L}}_{\mathrm{dd}}^0+\eta_1 \bar{\bar{L}}_{\mathrm{dd}}^1 & -\eta_0 \bar{\bar{K}}_{\mathrm{dd}}^0-\eta_0 \bar{\bar{K}}_{\mathrm{dd}}^1 & \eta_0 \bar{\bar{L}}_{\mathrm{dg}}^0 \\
\eta_0 \bar{\bar{K}}_{\mathrm{dd}}^0+\eta_0 \bar{\bar{K}}_{\mathrm{dd}}^1 & \eta_0 \bar{\bar{L}}_{\mathrm{dd}}^0+\frac{\eta_0^2}{\eta_1} \bar{\bar{L}}_{\mathrm{dd}}^1 & \eta_0 \bar{\bar{K}}_{\mathrm{dg}}^0+\frac{\eta_0}{2} \bar{\bar{S}}_{\mathrm{dg}} \\
\eta_0 \bar{\bar{L}}_{\mathrm{gd}}^0 & -\eta_0 \bar{\bar{K}}_{\mathrm{gd}}^0+\frac{\eta_0}{2} \bar{\bar{S}}_{\mathrm{gd}} & \eta_0 \bar{\bar{L}}_{\mathrm{gg}}^0
\end{bmatrix} \\
& \bar{\bar{Z}}_{12}=\begin{bmatrix}
\big(\eta_0 \bar{\bar{L}}_{\mathrm{dr}}^0\big)^{\mathrm{T}} & \big(\eta_0 \bar{\bar{K}}_{\mathrm{dr}}^0+\frac{\eta_0}{2} \bar{\bar{S}}_{\mathrm{dr}}\big)^{\mathrm{T}} & \big(\eta_0 \bar{\bar{L}}_{\mathrm{gr}}^0\big)^{\mathrm{T}}
\end{bmatrix}^{\mathrm{T}} \\
& \bar{\bar{Z}}_{21}=\begin{bmatrix}
\eta_0 \bar{\bar{L}}_{\mathrm{rd}}^0 & -\eta_0 \bar{\bar{K}}_{\mathrm{rd}}^0+\frac{\eta_0}{2} \bar{\bar{S}}_{\mathrm{rd}} & \eta_0 \bar{\bar{L}}_{\mathrm{gr}}^0
\end{bmatrix} \\
& \bar{\bar{Z}}_{22}=\eta_0 \bar{\bar{L}}_{\mathrm{rr}}^0.
\end{aligned}
\label{eq:19}
\end{equation}
To carry out the sub-structure CMA, first, the right hand side $\bar{V}_{\mathrm{1}}$ of~\eqref{eq:18} is set to zero, then the first row is inverted for $\bar{I}_1$, and the result is inserted into the second row. This yields
\begin{equation}
\underbrace{\big(\bar{\bar{Z}}_{22}-\bar{\bar{Z}}_{21} \big(\bar{\bar{Z}}_{\mathrm{11}}\big)^{-1} \bar{\bar{Z}}_{12}\big)}_{\displaystyle\bar{\bar{Z}}_{\mathrm{sub}}}\bar{I}_2=\bar{V}_{\mathrm{2}}
\label{eq:20}
\end{equation}
where $\bar{\bar{Z}}_{21} \big(\bar{\bar{Z}}_{\mathrm{11}}\big)^{-1} \bar{\bar{Z}}_{12}$ is the perturbation matrix. This term is also referred to as the numerical Green function matrix and represents the electromagnetic coupling effects from the dielectric substrate and the ground plane to the radiation patch.

The generalized eigenvalue equation corresponding to the linear system in~\eqref{eq:20} is given as~\cite{subJ2012},
\begin{align}
{\bar{\bar{X}}_{{\mathrm{sub}}}} {\bar{I}_n} = {\lambda _n}{\bar{\bar{R}}_{{\mathrm{sub}}}} {\bar{I}_n}
\end{align}
where $\lambda _n$ and $\bar{I}_n$ are the $n$th eigenvalue and corresponding eigenvector, and $\bar{\bar{X}}_{{\mathrm{sub}}}$ and $\bar{\bar{R}}_{{\mathrm{sub}}}$ are the imaginary and the real parts of $\bar{\bar{Z}}_{{\mathrm{sub}}}$, respectively.

\section{Numerical Results}

In this section, three different numerical examples that demonstrate the applicability and the accuracy of the proposed CM method are presented. In all examples, CMA is carried out at frequency samples in a given frequency range, and the resonance frequency is identified as the frequency sample where the eigenvalue $\lambda_n$ of a mode is closest to zero. This is because eigenvalues may not be exactly zero due to the sampling in frequency and the numerical errors in constructing the matrix system in~\eqref{eq:18}. For each example, the modal significance (MS) is computed using $\mathrm{MS }= 1 /\left|1+j \lambda_n\right|$~\cite{CM1971Harrington} and is plotted against frequency. Results obtained using (i) the full-structure CM method with global MTF, (ii) the sub-structure CM method with global MTF, (iii) the full-structure CM method with VSIE (as implemented by the commercial software FEKO), (iv) the sub-structure CM method with VSIE~\cite{huang2021accuratem,VSIEwuqi}, (v) the full-structure CM method with EFIE-PMCHWT~\cite{fan2022spurious}, and (vi) the CM method with MPIE~\cite{CM2015} are compared for different examples considered in this section.

\subsection{Rectangular Patch Structure}

In the first numerical example, the rectangular PEC patch of the antenna entirely covers the upper surface of an FR-4 substrate with the length $100\,\mathrm{mm}$, width $40\,\mathrm{mm}$, height $1.55\,\mathrm{mm}$.  The substrate is non-magnetic and its relative permittivity is $4.7$. The frequency band ranges from $1.0\,\mathrm{GHz}$ to $3.0\,\mathrm{GHz}$ and is sampled with a step of $25\,\mathrm{MHz}$. The average edge length of the triangular elements discretizing the surfaces of the antenna is $4.0\,\mathrm{mm}$.

\begin{figure}[t]
\centering
\subfigure[]{\includegraphics[width=0.49\columnwidth,draft=false]{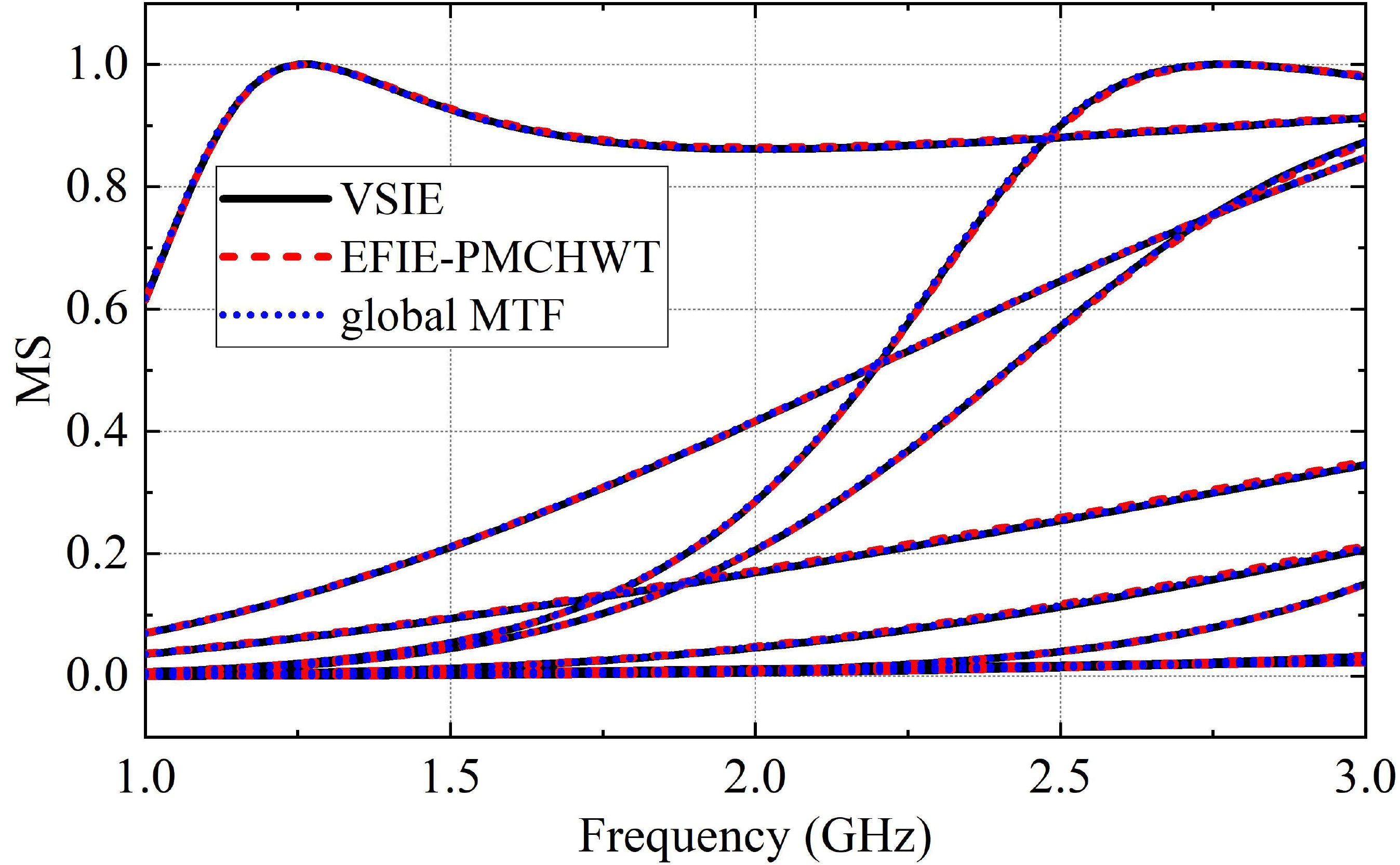}}
\subfigure[]{\includegraphics[width=0.49\columnwidth,draft=false]{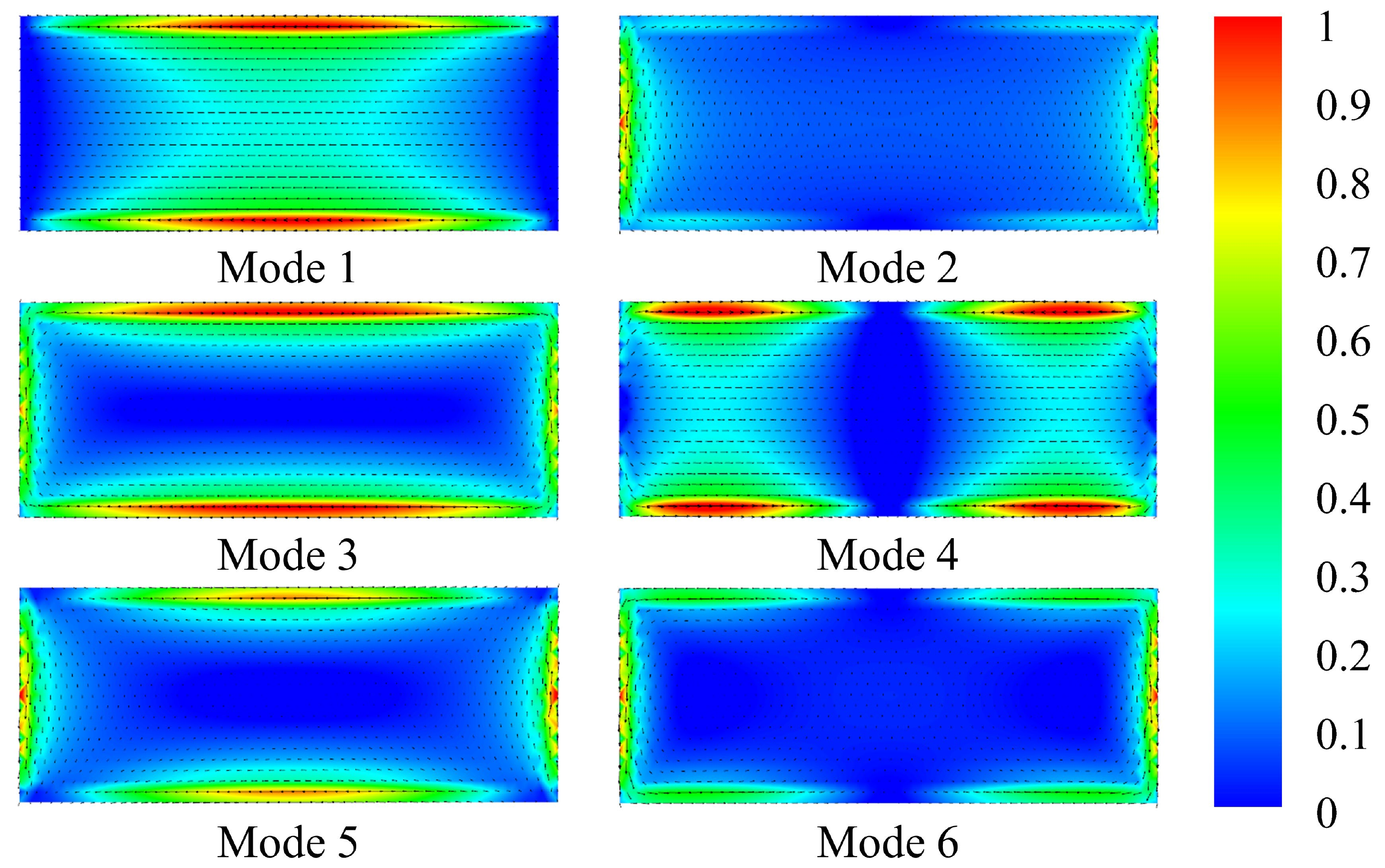}}
\caption{(a) The MS curves of the first 9 full-structure CMs of the rectangular patch antenna. (b) Amplitude of the first 6 electric eigencurrents obtained by the proposed sub-structure CM method with global MTF on the PEC patch at $1.275\,\mathrm{GHz}$.}
\vspace{-0.4cm}
\label{fig:PatFullMS}
\end{figure}

First, Fig.~\ref{fig:PatFullMS}(a) plots MS computed using the full-structure CM methods with global MTF, VSIE, and EFIE-PMCHWT versus frequency. When computing the full-structure CMs, the radiation weighting matrices of generalized eigenvalue equation are chosen following the principle described in~\cite{yla2019generalized} to avoid the contamination of the nonphysical modes. Clearly, all three results agree well.
The resonance frequencies computed by the full-structure CM methods with VSIE and global MTF are both $1.250\,\mathrm{GHz}$ and $2.775\,\mathrm{GHz}$, which verifies the accuracy of global MTF in full-structure CMA.

Next, the accuracy of the sub-structure CM method with global MTF is studied. The first resonance frequency is computed as $1.275\,\mathrm{GHz}$, which is the same value obtained by the sub-structure CM method with VSIE~\cite{VSIEwuqi}. Fig.~\ref{fig:PatFullMS}(b) plots the amplitude of the first $6$ electric eigencurrents on the PEC patch of the antenna at $1.275\,\mathrm{GHz}$ as obtained by the proposed method. These match the results presented in~\cite{VSIEwuqi}.

\subsection{Triangular Microstrip Patch antenna}

\begin{figure}[t]
\centering
\subfigure[]{\includegraphics[width=0.6\columnwidth,draft=false]{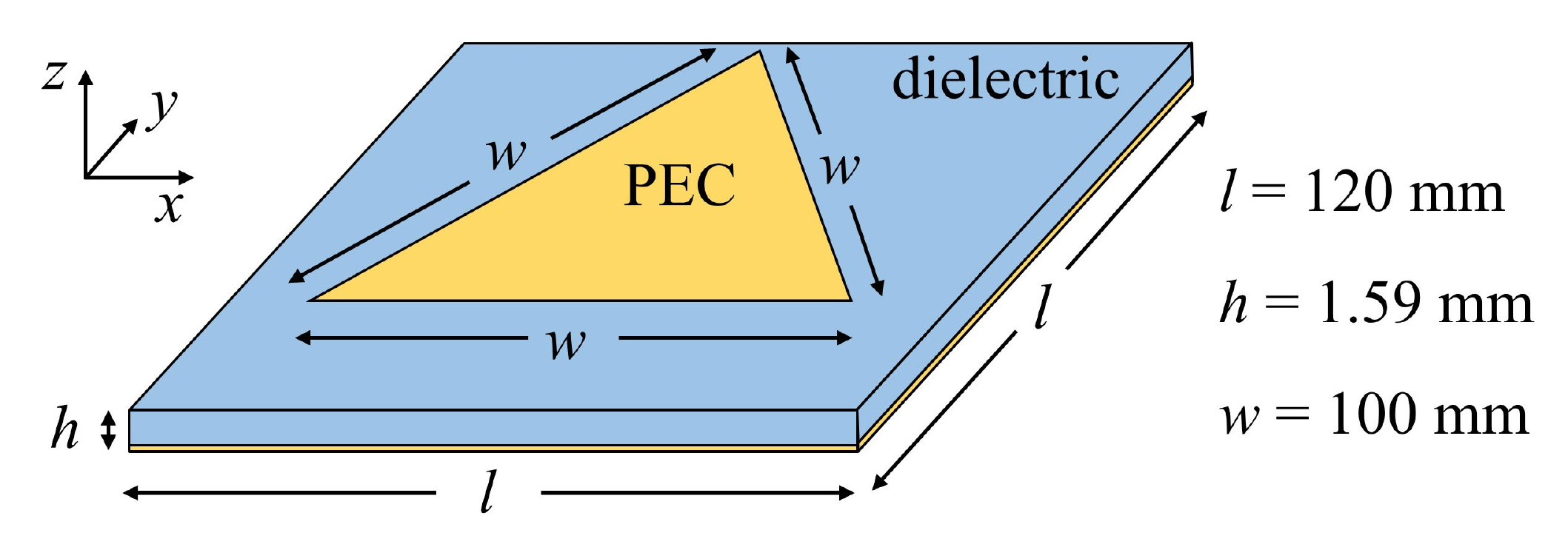}}
\subfigure[]{\includegraphics[width=0.49\columnwidth,draft=false]{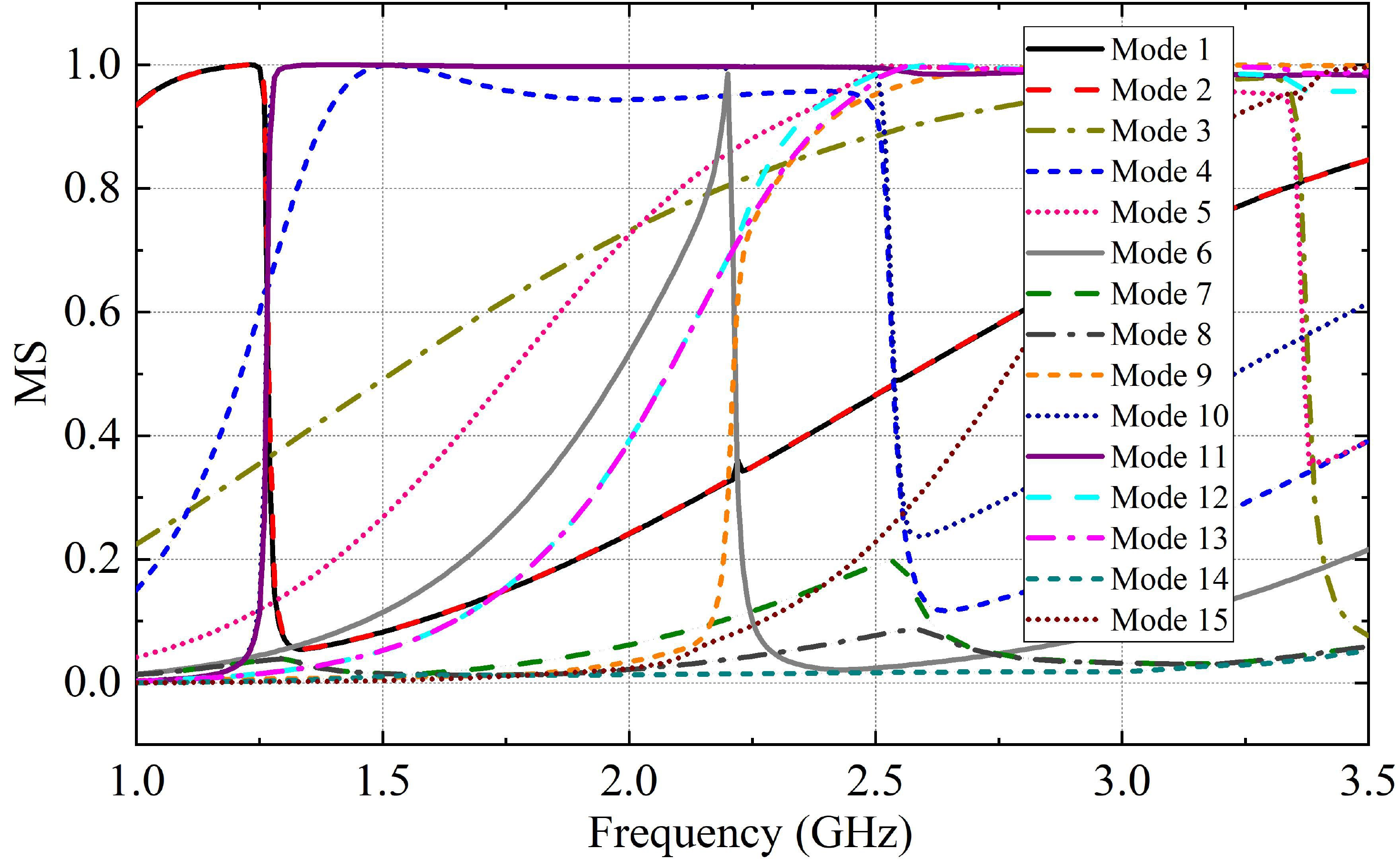}}
\subfigure[]{\includegraphics[width=0.49\columnwidth,draft=false]{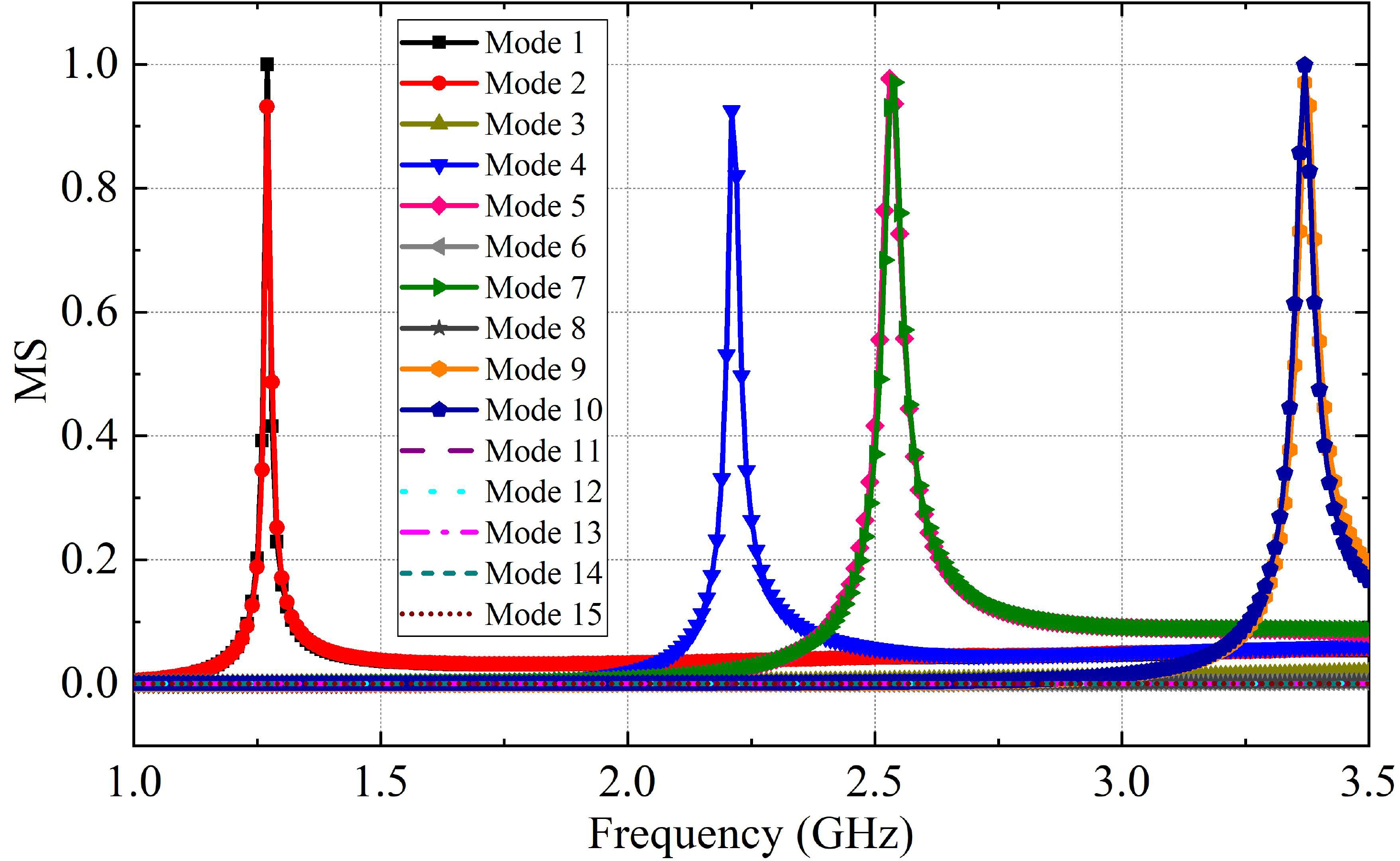}}
\caption{(a) Geometry of the triangular microstrip patch antenna. (b) The MS curves of the first $15$ CMs obtained using the full-structure CM method with EFIE-PMCHWT. (c) The MS curves of the first $15$ CMs obtained using the proposed sub-structure CM method with global MTF.}
\vspace{-0.2cm}
\label{fig:triaCM}
\end{figure}

In this example, the antenna consists of a triangular PEC patch, a dielectric substrate, a PEC ground plane as shown in Fig.~\ref{fig:triaCM}(a). The center of the triangular patch coincides with that of the upper surface of the substrate. The dimensions of the geometry are provided on the figure. The substrate is non-magnetic and its relative permittivity is $2.32$. The frequency band ranges from $1.0\,\mathrm{GHz}$ to $3.5\,\mathrm{GHz}$ and is sampled with a step of $10\,\mathrm{MHz}$. The average edge length of the triangular elements discretizing the surfaces of the antenna is $5\,\mathrm{mm}$.

Fig.~\ref{fig:triaCM}(b) and Fig.~\ref{fig:triaCM}(c) plot MS for the first $15$ CMs computed by the full-structure CM method with EFIE-PMCHWT and the sub-structure CM method with global MTF, respectively. Comparing these two figures shows that the full-structure CM method produces MS curves with wider bands, which might not be directly useful since the microstrip antenna design often yields narrow-band operation in the vicinity of a resonant frequency~\cite{huang2021accuratem}.

Table~\ref{t:triafreq} provides the first $7$ resonance frequencies obtained via measurements~\cite{triaC1984}, using the cavity model~\cite{triaJ1988}, the CM method with MPIE~\cite{CM2015}, the sub-structure CM method with global MTF, and the full-structure CM method with EFIE-PMCHWT. The table shows that the resonance frequencies obtained using the proposed sub-structure CM method with global MTF are in agreement with those obtained using the first three approaches.

Note that difference in the results obtained by the sub-structure CM method with global MTF and the full-structure CM method with EFIE-PMCHWT can simply be explained by the fact that full- and sub-structure methods produce different sets of modes. The results obtained using a sub-structure CM method are often more useful in the analysis and the design of a microstrip antenna since these results describe the modal behavior of the electromagnetic fields on the radiation patch and the feeding region.

\begin{figure}[!t]
\centering
\subfigure[]{\includegraphics[width=0.56\columnwidth,draft=false]{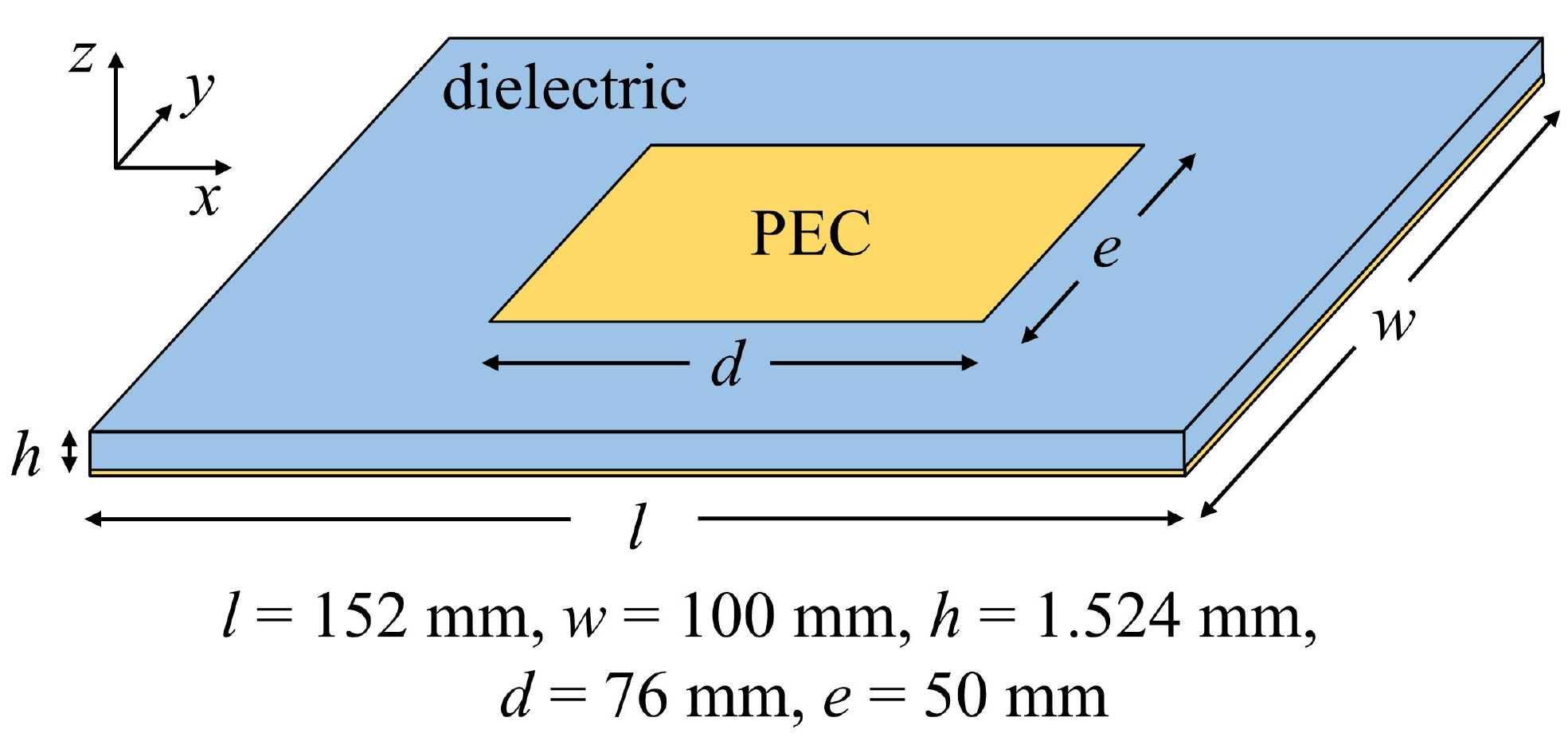}}\\
\subfigure[]{\includegraphics[width=0.49\columnwidth,draft=false]{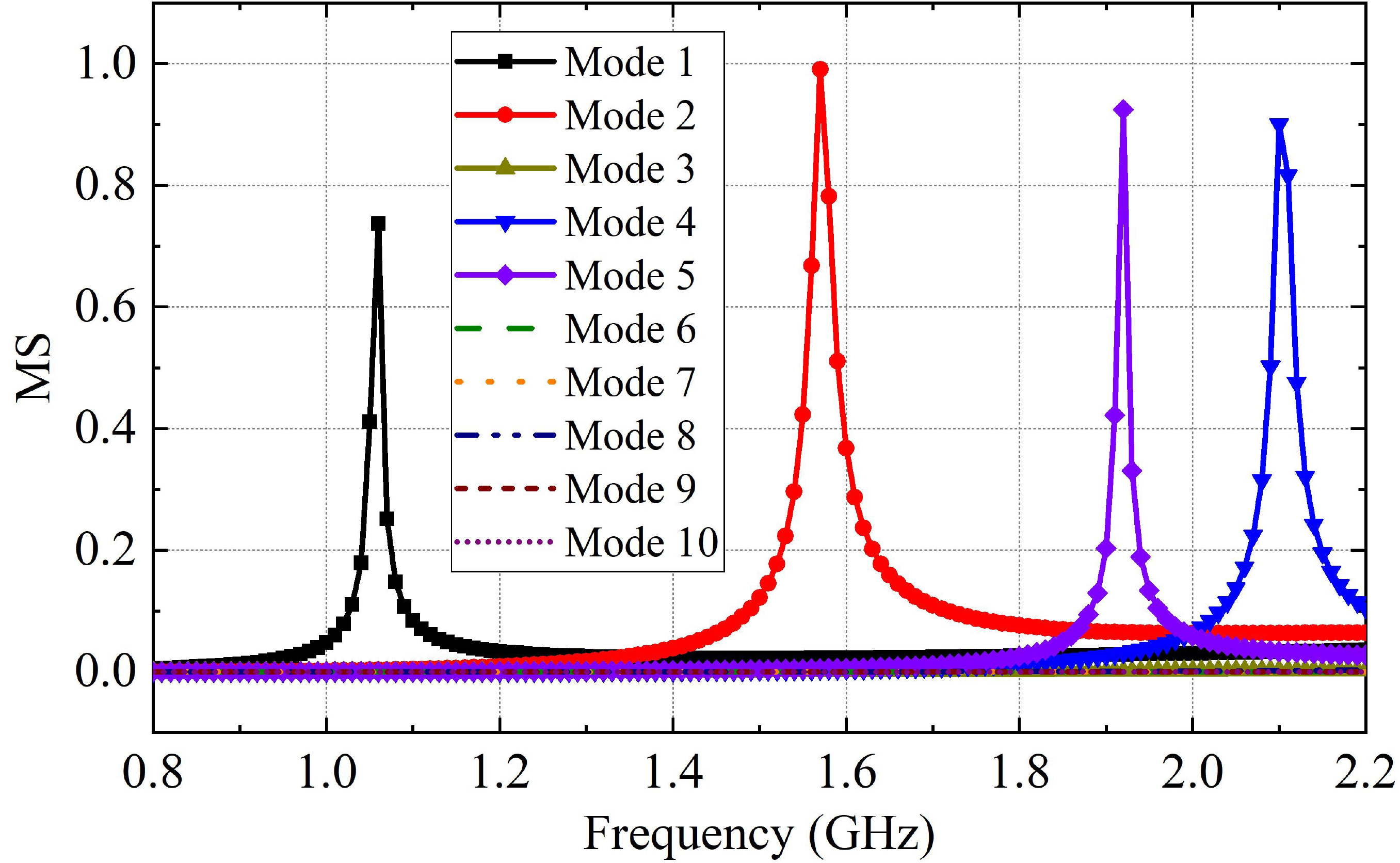}}
\subfigure[]{\includegraphics[width=0.49\columnwidth,draft=false]{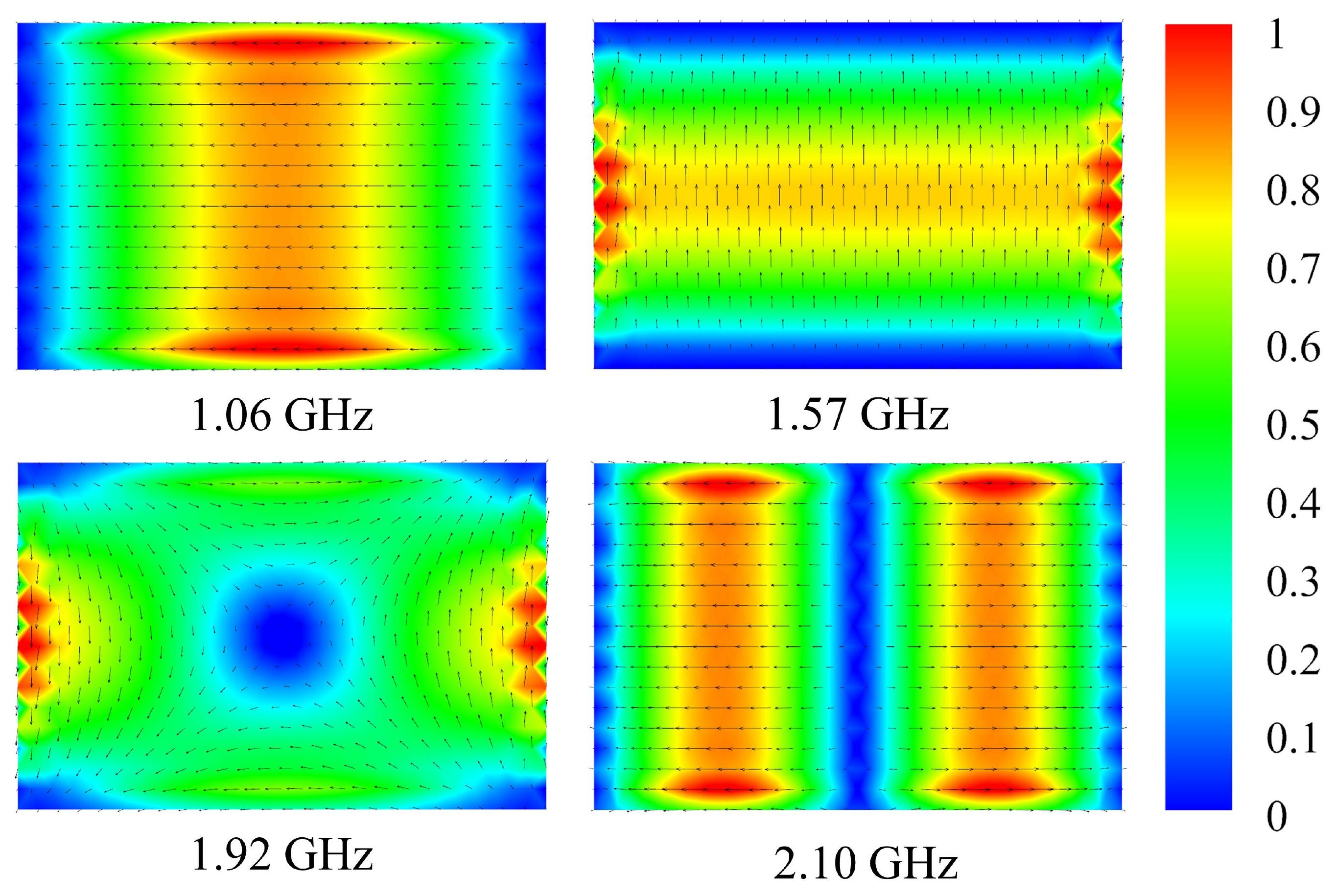}}
\subfigure[]{\includegraphics[width=0.49\columnwidth,draft=false]{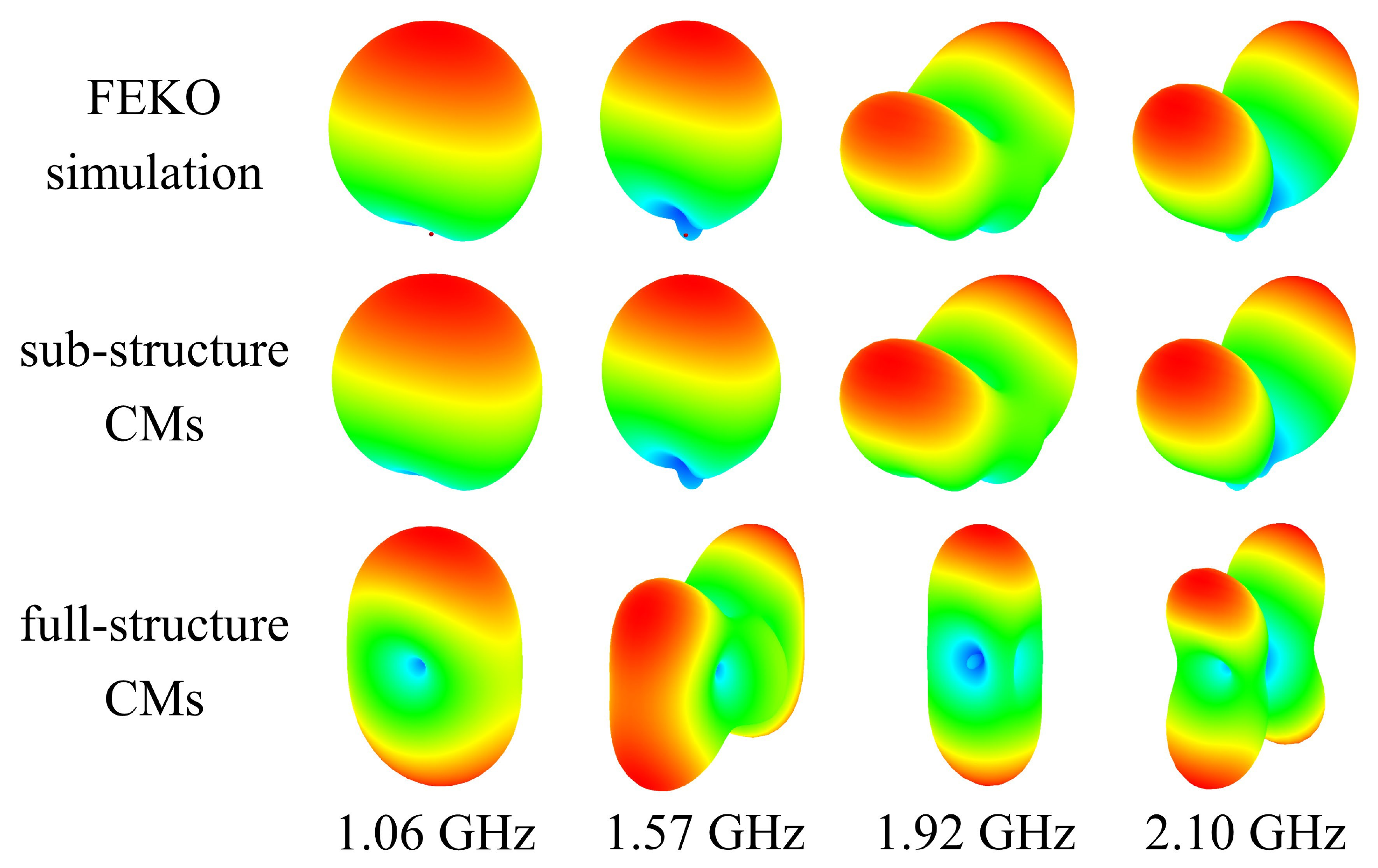}}
\caption{(a) Geometry of the rectangular microstrip patch antenna. (b) The MS curves of the first $10$ CMs obtained using the proposed sub-structure CM method with global MTF. (c) Amplitude of the electric eigencurrents obtained using the proposed sub-structure CM method with global MTF on the PEC patch at resonance frequencies $1.06\,\mathrm{GHz}$, $1.57\,\mathrm{GHz}$, $1.92\,\mathrm{GHz}$, and $2.10\,\mathrm{GHz}$. (d) Radiation patterns obtained using FEKO and characteristic fields obtained using the proposed sub-structure CM method with global MTF and the full-structure CM method with EFIE-PMCHWT at $1.06\,\mathrm{GHz}$, $1.57\,\mathrm{GHz}$, $1.92\,\mathrm{GHz}$, and $2.10\,\mathrm{GHz}$.}
\vspace{-0.5cm}
\label{fig:rectCM}
\end{figure}

\begin{table}[!t]
\centering
\caption{Resonance frequencies (GHz) of the triangular microstrip patch antenna}
\label{t:triafreq}
\renewcommand{\arraystretch}{1.05}
{\begin{tabular}{ c | c | c | c | c }
\makecell[c]{Measured}
&\makecell[c]{Cavity\\model}
&\makecell[c]{CM method\\with MPIE}
&\makecell[c]{CM method\\with MTF}
&\makecell[c]{CM method with\\EFIE-PMCHWT}\\ \hline
$1.28$& $1.30$& $1.30$& $1.27$& $1.23$\\ \hline
$1.28$& $1.30$& $1.30$& $1.27$& $1.23$\\ \hline
$2.24$& $2.25$& $2.25$& $2.21$& $1.40$\\ \hline
$2.55$& $2.60$& $2.57$& $2.53$& $1.40$\\ \hline
$2.55$& $2.60$& $2.57$& $2.54$& $1.52$\\ \hline
$3.40$& $3.44$& $3.44$& $3.37$& $2.20$\\ \hline
$3.40$& $3.44$& $3.44$& $3.37$& $2.51$\\
\end{tabular}}
\end{table}

\begin{table}[t]
\centering
\caption{Resonance frequencies (GHz) of the rectangular microstrip patch antenna}
\label{t:rectfreq}
\renewcommand{\arraystretch}{1.05}
{\begin{tabular}{c|c|c|c|c}
\makecell[c]{Simulated \\ $S_{11}$}
&\makecell[c]{Cavity \\model}
&\makecell[c]{CM method \\ with MPIE}
&\makecell[c]{CM method \\ with VSIE}
&\makecell[c]{CM method \\ with MTF}\\ \hline
$1.05$& $1.08$& $1.08$& $1.09$&  $1.06$\\ \hline
$1.55$& $1.61$& $1.60$& $1.65$&  $1.57$\\ \hline
$1.90$& $1.96$& $1.95$& $2.03$&  $1.92$\\ \hline
$2.10$& $2.15$& $2.13$& $2.19$&  $2.10$\\
\end{tabular}}
\end{table}

\subsection{Rectangular Microstrip Patch Antenna}

In the last example, the antenna consists of a rectangular PEC patch, a dielectric substrate, and a PEC ground plane as shown in Fig.~\ref{fig:rectCM}(a). The center of rectangular patch coincides with that of the upper surface of the substrate. The dimensions of the geometry are provided on the figure. The substrate is non-magnetic and its relative permittivity is $3.38$. The frequency band ranges from $0.8\,\mathrm{GHz}$ to $2.2\,\mathrm{GHz}$, and is sampled with a step of $10\,\mathrm{MHz}$. The average edge length of the triangular elements discretizing the surfaces of the antenna is $6\,\mathrm{mm}$.

Fig.~\ref{fig:rectCM}(b) plots MS of the first $10$ CMs computed using the sub-structure CM method with global MTF versus frequency. The resonance frequencies can clearly be identified. Table~\ref{t:rectfreq} compares the first 4 resonance frequencies computed using the sub-structure CM method with global MTF to those obtained using the simulated $S_{11}$ data in~\cite{CM2015}, the cavity model~\cite{BPA2006}, the CM method with MPIE~\cite{CM2015}, and the sub-structure method with VSIE~\cite{huang2021accuratem}. All results agree well with each other.

Fig.~\ref{fig:rectCM}(c) plots the amplitude of the electric eigencurrents computed by the sub-structure CM method with global MTF on the PEC patch of the antenna at resonance frequencies $1.06\,\mathrm{GHz}$, $1.57\,\mathrm{GHz}$, $1.92\,\mathrm{GHz}$, and $2.10\,\mathrm{GHz}$. These results are in agreement with those obtained using the CM method with MPIE.

Furthermore, the radiation patterns of the antenna are computed using FEKO. The rectangular patch is fed by a port source as described in~\cite{CM2015} to excite the above 4 modes. Fig.~\ref{fig:rectCM}(d) plots the radiation patterns obtained using FEKO and characteristic fields obtained using the sub-structure CM method with global MTF and the full-structure CM method with EFIE-PMCHWT
at $1.06\,\mathrm{GHz}$, $1.57\,\mathrm{GHz}$, $1.92\,\mathrm{GHz}$, and $2.10\,\mathrm{GHz}$. The figure shows that the simulated radiation patterns agree well with the characteristic fields obtained using the sub-structure CM method, demonstrating that the sub-structure methods are often more suitable to be used in the design of microstrip antennas.

\section{Conclusion}
In this work, a sub-structure CM method that relies on the global MTF of SIEs is formulated and implemented to compute the modes and the resonance frequencies of practical microstrip patch antennas with finite dielectric substrates and ground planes. The global MTF naturally separates the equivalent currents on the radiation patch from those on the surface of the dielectric substrate, which yields a matrix that can be easily put in a form that is suitable for sub-structure CM analysis. The resulting sub-structure CM method avoids the cumbersome computation of the multilayered medium Green function (unlike the CM methods that rely on MPIE) and the volumetric discretization of the substrate (unlike the CM methods that rely on VSIE), and numerical results show that it is a reliable and accurate approach to predicting the modal behavior of electromagnetic fields on practical microstrip antennas. An extension of the proposed method, which can account for lossy dielectric substrates, is underway.

  \newpage

%


%

\bibliographystyle{IEEEtran}
\bibliography{References}

\end{document}